\documentclass[twocolumn,prb,aps,superscriptaddress]{revtex4}
\usepackage{color}
\usepackage{graphicx}
\usepackage{subfig}
\usepackage{natbib}
\usepackage[fleqn]{amsmath}
\usepackage{amssymb}
\newcommand{\nc}{\newcommand}
\nc{\on}{\operatorname}
\nc{\wt}{\widetilde}
\nc{\Wick}{{\mathbb :}}
\nc{\R}{{\mathbb R}}

\newcommand{\beq}{\begin{equation}}
\newcommand{\eeq}{\end{equation}}
\newcommand{\bmul}{\begin{multline}}
\newcommand{\emul}{{\end{multline}}}
\newcommand\beqa{\begin{eqnarray}}
\newcommand\eeqa{\end{eqnarray}}
\newcommand\bea{\begin{array}}
\newcommand\eea{\end{array}}
\newcommand\ba{\begin{array}}
\newcommand\ea{\end{array}}

\newcommand{\neqa}{\nonumber\end{eqnarray}}

\nc{\CH}{{\mathcal H}}
\nc{\Db}{{\bar D}}
\nc\comment[1]{}

\nc{\CM}{{\mathcal M}}
\nc{\CN}{{\mathcal N}}

\newcommand{\re}{\relax{\rm I\kern-.18em R}}

\nc{\meV}{{\mathrm{\,meV}}}
\nc{\cG}{{\mathcal G}}

\renewcommand{\)}{\right)}

\comment{ \)}
\renewcommand{\bar}{\overline}

\nc{\al}{{\alpha}}

\begin{document}

\title{Pechukas-Yukawa approach to the evolution of the quantum state of a parametrically perturbed system}
\author{Mumnuna A. Qureshi}
\address{Department of Physics, Loughborough University, Loughborough LE11 3TU, UK.}
\author{Johnny Zhong}
\address{Department of Mathematical Sciences, Loughborough University, Loughborough LE11 3TU, UK.}
\author{Zihad Qureshi}
\address{Attero Solutions, Chatham, UK.}
\author{Peter Mason }
\address{Department of Physics, Loughborough University, Loughborough LE11 3TU, UK.}
\author{Joseph J. Betouras }
\address{Department of Physics, Loughborough University, Loughborough LE11 3TU, UK.}
\author{Alexandre M. Zagoskin}
\address{Department of Physics, Loughborough University, Loughborough LE11 3TU, UK.}

\begin{abstract}
We consider the evolution of the quantum states of a Hamiltonian that is parametrically perturbed via a term proportional to the adiabatic parameter $\lambda(t)$. 
Starting with the Pechukas-Yukawa mapping of the energy eigenvalues evolution on a generalised Calogero-Sutherland model of 1D classical gas, we consider the adiabatic approximation with two different expansions of the quantum state in powers of $d\lambda/dt$ and compare them with a direct numerical simulation. We show that one of these expansions (Magnus series) is especially convenient for the description of non-adiabatic evolution of the system. Applying the expansion to the exact cover 3-satisfiability problem, we obtain the occupation dynamics which provides insight on the population of states and sources of decoherence in a quantum system.
\end{abstract}
\maketitle

\section{Introduction}

Quantum computers offer significant advantages over their classical counterparts. The solution is encoded in the final quantum state of the system, which can be reached only through a series of highly entangled intermediate states. This is a formidable task given the intrinsic vulnerability to decoherence. In light of the considerable challenges posed in controlling and manipulating a large register of qubits, an alternative protocol, adiabatic quantum computing (AQC), has been proposed\cite{Fahri1, Fahri2}. In such a protocol, the system is initially prepared in an easily achievable ground state; provided the adiabatic evolution is sufficiently well-controlled, the end-state will be the ground state solution of the desired problem.

One can approach this problem through an analysis of the energy spectrum of the system, given by the following Hamiltonian:

\begin{equation}
\begin{gathered}
\label{GenHam}
H(\lambda(t)) = H_0 + \lambda(t)ZH_b,
\end{gathered}
\end{equation}
where $H_0$ is an unperturbed Hamiltonian with an easily achievable nondegenerate ground state, $\lambda(t)$ is an adiabatically evolving parameter and the perturbation $ZH_b$ is a large bias term with $Z \gg 1$\cite{Zagoskin1, Wilson1, Wilson2}. 

Pechukas\cite{Pechukas} and Yukawa\cite{Yukawa3} developed a formalism, mapping the level dynamics of Eq.(\ref{GenHam}) to a one-dimensional (1D) classical gas with inverse cubic repulsion. Remarkably, under this exact mapping, the entire Hamiltonian dynamics are determined by the initial conditions of the gas particles, and thus the solution of the problem is encoded in classical initial conditions. The evolution of the energy spectrum then provides useful information on the evolution of the energy gap and the distribution of avoided crossings. Using this description, we connect the level dynamics of a system to the quantum states\cite{Ours} through the  evolution of $C(t)$, for a wavefunction expanded in the instantaneous eigenstates $\psi=\sum_n{C_n(t)|n\rangle}$. The equilibrium statistical mechanics of the Pechukas-Yukawa `gas' were instrumental in obtaining the results of random matrix theory\cite{Haake}. The advantage under this decription for a quantum coherent system is that the instantaneous eigenstates include all higher level entanglements.

One can extend this description from eigenvalue dynamics to determine the form of the density matrices. This provides insight in the dynamics of occupation numbers and the coherences in the system which will prove useful in determining the probability for the system to remain in its initial state. Using this description, one can, for example, determine the effects of avoided level crossings on the system's evolution and the extent to which the noise affects the population of states.  In an earlier paper \cite{Ours}, we developed a consistent nonequilibrium formalism for this `gas' (the BBGKY chain), with the expectation to  apply it to the statistical analysis of classes of problems tractable (or not) by an AQC. The present work builds on [\onlinecite{Ours}], to further extend the model from the statistical mechanics of energy levels to the description of quantum states themselves. It is worth stressing that these works build a general scheme applicable to the investigation of AQC, however, they are not restricted to AQC.

To proceed we use a Magnus series expansion to approximate $C(t)$, a convenient way to obtain an assymptotic expansion. This approach is contrasted against both the adiabatic approximation and the time dependent perturbation theory (TDPT). We determine the coefficients of the eigenstates to compare how well these approximations accommodate adiabatic parameters\cite{Kato, Gernot}. Using the Magnus series, $C(t)$ can be approximated by a cumulant expansion to re-sum the TDPT, in powers of $\dot\lambda=d\lambda/dt$, with respect to the adiabaticity. Each term of the expansion corresponds to a sum of an infinite number of terms in a direct expansion of the density matrix. Given the Magnus series converges, the cumulant expansion provides a source of improved efficiency in the result. This is important to study the adiabatic invariants of the system. Knowledge of this could yield important features of the behaviour of an AQC.

Our analysis shows that the convergence of the Magnus series approximating the evolution of $C(t)$ is governed by the initial conditions. This could provide better insight into what measurable characteristics of a system can be used as a criterion for its quantum performance. This carries the potential to specify Hamiltonians of different complexity classes, governed by the initial conditions in the Pechukas-Yukawa formalism. Moreover it may be possible to extend the argument to stoquastic (stochastic quantum) systems where noise is added; this may prove crucial experimentally. 

The structure of the paper is as follows: In Sec. II we provide an overview of the Pechukas formalism and the evolution of the eigenstate coefficients before presenting details on the main result of our paper, in Sec. III, on the Magnus series approximation that we develop to study the evolution of the perturbed quantum system. The Magnus series is compared numerically against two other approximations; the adiabatic approximation and the time dependent perturbation theory (TDPT), investigating its limitations. These results are numerically tested by use of an example, determining the occupation dynamics numerically for the exact cover 3 NP-complete problem in Sec. IV. We discuss and conclude our work in Sec. V.  Furthermore we include two appendices that provide additional technical details.

\section{The Pechukas Model and the Evolution of eigenstate coefficients}

The Pechukas-Yukawa approach maps from quantum systems described by Eq.(\ref{GenHam}) to a classical set of Hamilton's equations. The level dynamics of the quantum system is modelled as classical fictitious gas particles moving in 1D with parametric evolution in time through $\lambda$; the number of fictitious particles $N$, corresponding to the number of levels in the classical Hamiltonian concerning ``position'' $x_n$, ``velocity" $v_n$ and particle-particle repulsion, analogous to ``relative angular momentum" $l_{mn}$. The dynamics is fully integrable, well suited though not restricted to adiabatic systems\cite{Haake, Zagoskin, Ours}. 

The fictitious particles interact via a pairwise repulsive potential, with associated Hamiltonian given by the following:

\begin{equation}
\begin{gathered}
\label{HamiltonianP}
H=\frac{1}{2}\sum^N_{n=1}v^2_n+\frac{1}{2}\sum^N_{n \neq m}\frac{|l_{mn}|^2}{(x_m-x_n)^2}.
\end{gathered}
\end{equation}

\noindent We assume the energy spectrum is non-degenerate such that as $\lambda$ varies, any accidental degeneracies are broken\cite{Pechukas}. The level dynamics of this system is governed by the following closed set of ordinary differential equations\cite{Pechukas, Yukawa1}: 

\begin{equation}
\begin{gathered}
\label{Pechukas}
\frac{dx_m}{d\lambda}=v_m, \\
\frac{dv_m}{d\lambda}=2\sum_{m\neq n}{\frac{{{|l}_{mn}|}^2}{{(x_m-x_n)}^3}},\\
\frac{d{l}_{mn}}{d\lambda}=\sum_{k\neq m,n}{l_{mk}l_{kn}\left(\frac{1}{{(x_m-x_k)}^2}-\frac{1}{({x_k-x_n)}^2}\right)},
\end{gathered}
\end{equation}

\noindent where $x_m\left(\lambda \right)=E_m(\lambda )=\left\langle m|H|m\right\rangle$, denoting the instantaneous eigenvalues of the system, $v_m\left(\lambda \right)=\left\langle m|ZH_b|m\right\rangle $ and $l_{mn}$ is defined by $l_{mn}\left(\lambda \right)=\left(E_m\left(\lambda \right)-E_n(\lambda )\right)\left\langle m|ZH_b|n\right\rangle $ satisifying the relation, $l_{mn}=-l^*_{nm}$. Each eigenvalue moves with a different ``velocity" in accordance with Eq. (\ref{Pechukas}) as $\lambda$ varies. Typically the diagonal elements of the potential are very different as the states have very different spatial distributions, therefore sample different regions of the potential\cite{Pechukas}. As a consequence of the coupling strengths between particle pairs becoming dynamic variables, the phase space of the system is greater than $2N$, where $N$ denotes the number of levels. In the special case when $l_{mn}$ is constant, the system becomes the Calogero-Sutherland model\cite{Haake, Chowski}. The Pechukas-Yukawa model is fully integrable and thus promises to lead to constants of motion in the system\cite{Stock}. 

This set of differential equations describes the aforementioned mapping of the level dynamics to that of a 1D classical gas\cite{Haake, Zagoskin}. The mapping of Eq. (\ref{GenHam}) to Eq. (\ref{Pechukas}) is an identical operation valid for an \textit{arbitrary} time dependent $\lambda$\cite{Pechukas, Haake, Ours}. Note that time does not explicitly enter Eq.(\ref{Pechukas}), the levels evolve parametrically in time through $\lambda$ which determines the instantaneous energy levels: this is a set of equations for the Hamiltonian, and not for (time- and initial state-dependent) quantum states of a system described by such a Hamiltonian. 

Using the Pechukas-Yukawa model, a link has been established between the level dynamics and the evolution of the eigenstate expansion coefficients,  $C_n(t)$\cite{Ours}, which can be extended to the evolution of the quantum states. This provides a description of the eigenstate coefficients in terms of all higher level entanglements which may prove advantageous, however this investigation is beyond the scope of this paper. The eigenstate expansion coefficients have been shown to satisfy the following set of coupled differential equations\cite{Ours}:

\begin{equation}
\begin{gathered}
\label{EvolC}
i\dot{C_m}(t)-C_m(t)x_m=i\dot\lambda(t)\sum_{n\neq m}{C_n}(t)\frac{l_{mn}}{{(x_m-x_n)}^2}.
\end{gathered}
\end{equation}


\noindent We denote: 
\[X=\mathrm{diag}\left(x_1\dots x_n\right),\]
\[P=p_{mn}$ where $p_{mn}=\frac{l_{mn}}{{\left(x_m-x_n\right)}^2}$ and $p_{mm}=0,\]
\[C(t)={\left(C_1(t)\dots C_n(t)\right)}^T,\]
so that Eq.(\ref{EvolC}) can be written in the form
 
\begin{equation}
\begin{gathered}
\label{EvolCMatrix}
\frac{\partial }{\partial t}C(t)=A(t)C(t),
\end{gathered}
\end{equation}

\noindent where $A(t)=(-iX+\dot\lambda(t)P)$, at different time instances, does not commute with itself. In the present work, we investigate approximate methods to solve for $C(t)$, from which the occupation numbers are obtained, expanding on the model devloped in [\onlinecite{Ours}] from the statistical mechanics of level dynamics to the description of quantum states. 

\section{Magnus series approximation}

\subsection{Magnus series}
The Magnus series provides a solution to Eq.(\ref{EvolCMatrix}), taking into account the non-commutativity of $A(t)$\cite{Magnus, Blanes, Gantmacher, Oteo}. We begin by writing $C(t)$ in the form:

\begin{equation}
\begin{gathered}
\label{GenMag}
 C(t)=e^{\Omega(t)}C_0,\\
\Omega(t)=\sum^{\infty}_{k=1}\Omega_{k}(t),
\end{gathered}
\end{equation}

\noindent where $C_0=C(0)$ is the initial conditions for $C(t)$. Here $\Omega_k$ corresponds to the $k^{th}$ order term of the Baker-Campbell-Hausdorff (BCH) formula \cite{Magnus, Oteo} and is given as integrals of successive commutators. This can be used to construct an infinite hierarchy of $\dot\lambda$ terms from a cumulant expansion, which both improves the efficiency of the series and allows for the study of the adiabatic properties of the system related to $C(t)$. The first two terms of the series for $\Omega_k(t)$ read:

\begin{equation}
\begin{gathered}
\label{MagTerms}
\Omega_{1}(t)=\int^t_{0}{A(s)ds},\\
\Omega_{2}(t)=\frac{1}{2}\int^t_{0}{\int^s_{0}{[A(s),A(s')]ds'}ds}.\\
\end{gathered}
\end{equation}

Since the full Magnus series is not tractable, one resorts to a truncation, approximating the solution. In extension investigating  the asymptotic convergence of this series would be of interest in future research. In our subsequent analysis, we truncate the Magnus series to the $2^{nd}$ order and test it numerically.

\subsection{Convergence of the Magnus series}

\noindent In the Pechukas model, all information for the Hamiltonian dynamics is encoded in its initial conditions; we translate the conditions for convergence of the full Magnus series in terms of initial conditions. The Magnus series converges if\cite{Magnus, Blanes, Gantmacher, Oteo}:

\begin{equation}
\begin{gathered}
\label{MagConv}
\int^t_{0}||{A(s)||ds} < \pi.
\end{gathered}
\end{equation}

\noindent Using the triangle inequality and the expression for $A(t)$, it suffices to show that:

\begin{equation}
\begin{gathered}
\label{ConvPech}
\int^t_{0}{||X||ds} +\int^t_{0}{\dot\lambda(s)||P||ds} <\pi.
\end{gathered}
\end{equation}

To rewrite the first integral in Eq.(\ref{ConvPech}) in terms of initial conditions $x_n(0), v_n(0),  l_{mn}(0)$, we express the Pechukas equations Eq. (\ref{Pechukas}) in Lax formalism\cite{Light, YukawaLax, Chowski}. From this we describe the $||X||$ integral by the following  (for details, refer to Appendix A):

\begin{widetext}
\begin{equation}
\begin{gathered}
\label{ConvPechX}
\int^t_{0}{\sqrt{||X(0)||^2+\lambda(s)Tr(X(0)Q(0))+\lambda^2(s)||Q(0)||^2}ds} \\
\leq t||X(0)||+\sqrt{Tr(X(0)Q(0))}\int_0^t{\sqrt{\lambda(s)}ds}+||Q(0)||\int_0^t{|\lambda(s)|ds},
\end{gathered}
\end{equation}
\end{widetext}

\noindent where $Q$ is defined in Appendix A. Thereby the convergence of the first integral is reduced solely to the dependence of initial conditions and the time evolution of $\lambda$. This method however, is restricted to finite times such that initial conditions can be set to satisfy Eq. (\ref{ConvPech}). As $t \rightarrow \infty$, it is not possible to meet this convergence critera regardless of the restrictions on the initial conditions. 


Similarly, for $||P||$,  using that the square root of a sum is less than the sum of the square roots and interchanging the sum and integral using Tonelli's theorem, we can rewrite the second integral in Eq.(\ref{ConvPech}):

\begin{equation}
\begin{gathered}
\label{ConvPechP}
\int^t_{0}{\dot\lambda(s)||P||ds} \leq \sum_{m \neq n}\int^t_{0}{\dot\lambda(s)p_{mn}ds}. 
\end{gathered}
\end{equation}

\noindent Taylor expanding around the initial time for short time intervals, $p_{mn}$ is expressed in terms of initial conditions, $p_{mn} = p_{mn}(0) + \delta\lambda(s) {\dot{p}_{mn}(0})$ (using our definition of $p_{mn}$ from before) where $\delta\lambda(s)=(\lambda(s)-\lambda(0))$. Then Eq. (\ref{ConvPechP}) becomes:

\begin{equation}
\begin{gathered}
\label{ConvPechTaylor}
 \int^{t}_{0}{\dot \lambda(s)  (p_{mn}(0) + \delta\lambda \dot{p}_{mn}(0))  }  ds  \\
= \frac{\dot p_{mn}(0)}{2} (\lambda^2(t)  - \lambda^2(0))+\\
 \delta \lambda(t) (p_{mn}(0) - \lambda(0)\dot{ p}_{mn}(0)),
\end{gathered}
\end{equation}

\noindent where $\dot p_{mn}$ can be computed entirely from $x_m(0), v_m(0)$ and $l_{mn}(0)$.  We conclude that using (\ref{ConvPech}), (\ref{ConvPechX}) and (\ref{ConvPechTaylor}), the convergence of Magnus series is guaranteed and is expressed entirely through its parametric evolution, $\lambda$ and initial conditions. 

A potential source of divergence of the Magnus series involves level crossings; in the case of Landau-Zener transitions, the system is simplified to 2 levels with linear evolution in $\lambda$ hence $\dot\lambda$ is constant.  In Appendix B, we show that level crossings can be disregarded as they have zero measure. 
%
%

From these expressions, one can determine (from the initial conditions encoding the evolution of the system) when the convergence criterion outlined in Eq.(\ref{ConvPech}) are met.


\subsection{Numerically comparing the Magnus series against the adiabatic approximation and TDPT }

We compare numerically the Magnus series (up to its second order) against the adiabatic approximation, treating $\dot\lambda$ as negligible. Under the adiabatic approximation, $C(t)={e}^{-i\int^t_{0}{X(s)ds}}C_0$. Both these approximations are contrasted against the TDPT expanding $C(t)$ in powers of the interaction. The TDPT is useful for time-independent exactly solevable systems with an interaction to its environment described by a small perturbation\cite{TDPT}. Under this description, $C(t) \approx \sum^{\infty}_{i=0} C^i(t)$, where  $C^i(t) = \int^t_0 A(s)C^{i-1}(s)ds$, represent higher order corrections. These are obtained iteratively for 10 iterations. This solution breaks down for the TDPT when perturbations are large. To avoid this initially, levels are chosen with a minimum spacing of 0.01 and 0.05. This ensures that initial $||P||$ is not large as a consequence of level (avoided) crossings. Levels tend to diverge away from each other as the system evolves, hence $||P||$ decreases with time. However, it is unavoidable that for large $N$ level (avoided) crossings would not occur. This approach is sensitive to the time steps of evolution, requiring that they be small. The TDPT depends on the quantum states $C(t)$; unlike both the Magnus series and the adaiabtic approximation, where comparisons are made between matrix propagators in determining relative error. 

To compare these methods numerically, we take a piecewise constant approximation. Treating $A$ as constant over sufficiently small time steps, such that the TDPT is applicable, we break the interval of evolution in steps of 0.01. This approximation numerically converges to the true solution. This explicit  solution is given by:

\begin{equation}
\begin{gathered}
\label{PC}
C(t) = \prod_{i=0}{{e}^{(t_i - t_{i-1})A_i}}C_0,
\end{gathered}
\end{equation}

\noindent where $0 \leq t_0 < t_1 < \dots \leq t$ and $A_i$ is constant on interval $[t_{i-1},t_i]$.  


We investigate different classes of  Hamiltonians, each parameterised by their initial conditions $H(\lambda(t); x^0,v^0, l^0)$ with $ x^0,v^0$ and $l^0$ describing the initial time level dynamics, governed by functions of $\lambda$(t): 1) linear $\lambda(t) = 10^{-3}t$, 2) cubic $ \lambda(t) = 10^{-3}(t^3+t^2+t)$ and 3) exponential decay; $\lambda(t) = 10^{-3}e^{-t}$.  In accordance to Eq.(\ref{ConvPechX}) and Eq.(\ref{ConvPechTaylor}), the upper bound on the convergence criterion of the Magnus series  grows as $\mathcal{O}(t^2)$ for linear functions of $\lambda$, $\mathcal{O}(t^6)$ for the cubic function  and $\mathcal{O}(t)$ for the exponential decay. This suggests the convergences are expected to hold longest for an exponential decay. Under the same initial conditions, these different $\lambda$ yield the same level dynamics. Fig. 1 depicts the level dynamics for a linear function of $\lambda$.

\begin{figure}
\includegraphics[width=0.9\linewidth]{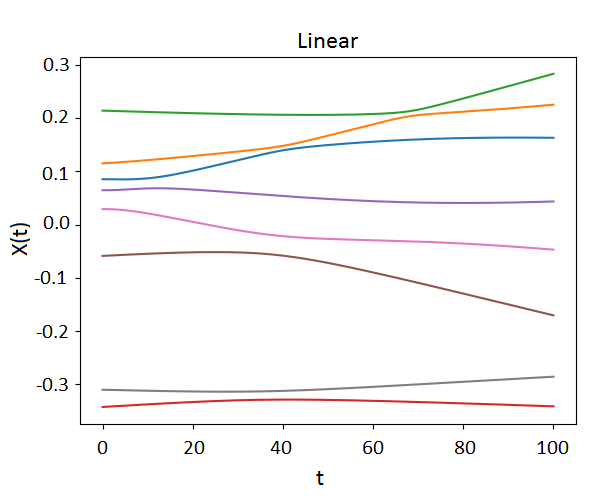}
 \caption{The time evolution of an 8 level system, for $t \in [0, 100]$. Levels have an initial minimum spacing of 0.05. The dynamics is encoded in the initial conditions, governed by $\lambda$ being the linear function of time.  Different $\lambda$ correspond to different nonlinear stretchings in the dynamics against time. Given the inital conditions are the same, the dynamics are the same.  We observe multiple avoided crossings between the different levels during their dynamics. We note that the levels are seen to be moving away from each other as time evolves. } 
\label{fig:Dynamics}
\end{figure}

We use the Euler method with random initial conditions uniformly distributed over a ball of radius $\frac{\pi}{6}$ to evolve the general Pechukas equations Eq. (\ref{Pechukas}), such that the conditions outlined in Eq.(\ref{ConvPech}) are met for 0-1 in steps of 0.01 for 1000 simulations to average over the random initial conditions for $x, v, l$. We evolve the dynamics up to t=100, without amending initial conditions in order to observe the limitations of the Magnus series. We compare the logarithm of the relative errors between the piecewise constant approach given by Eq.(\ref{PC}). The average relative error (R.E.), at each time step per simulation is given by $R.E.=\frac{1}{1000}\sum^{1000}_{i=1}\frac{||\tilde{C}[i]-C_{PC}[i]||}{||C_{PC}[i]||}$, where $\widetilde{C[i]}$ describes the approximation of $C(t)$ and $C_{PC}[i]$ the piecewise constant solution at time step $i$. Taking the norm provides a real valued relative error to plot against time. We take $N=2,4, 8, 12$ excited states for an initial minimum level spacing of 0.01 and $N=2,4,5$ excited states for 0.05 to check the effectiveness of the methods as the dimensionality increases. Note that for a radius of $\frac{\pi}{6}$ for the distribution of initial conditions, it is not possible for a minimum level spacing of 0.05 beyond $N=5$. Comparative results are given in Figs. 2, 3 and 4 for minimum initial level spacing 0.01 and Figs. 5, 6 and 7 for 0.05. These detail the growth of the logarithmic relative errors with time between the approximations, for each function of $\lambda$.

\begin{figure*}
\begin{center} 
\includegraphics[width=0.9\linewidth]{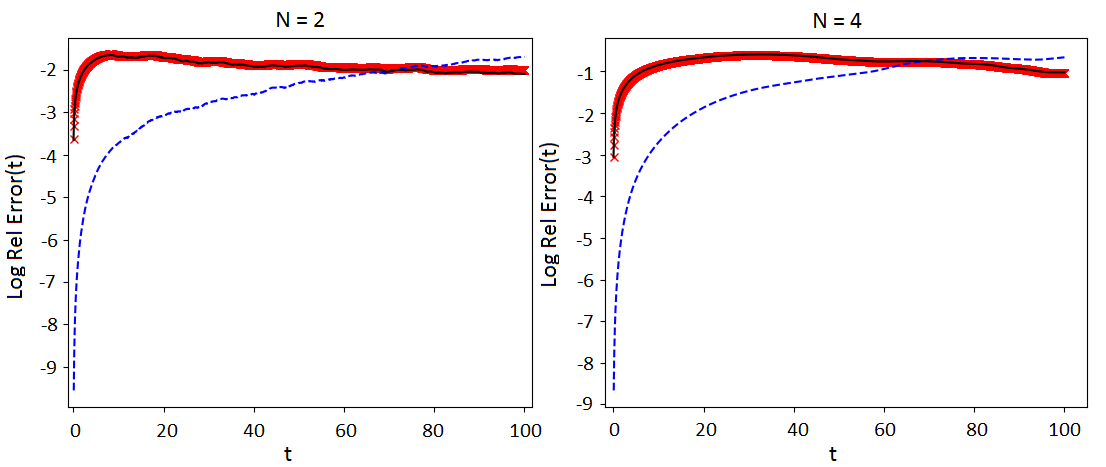}
\includegraphics[width=0.9\linewidth]{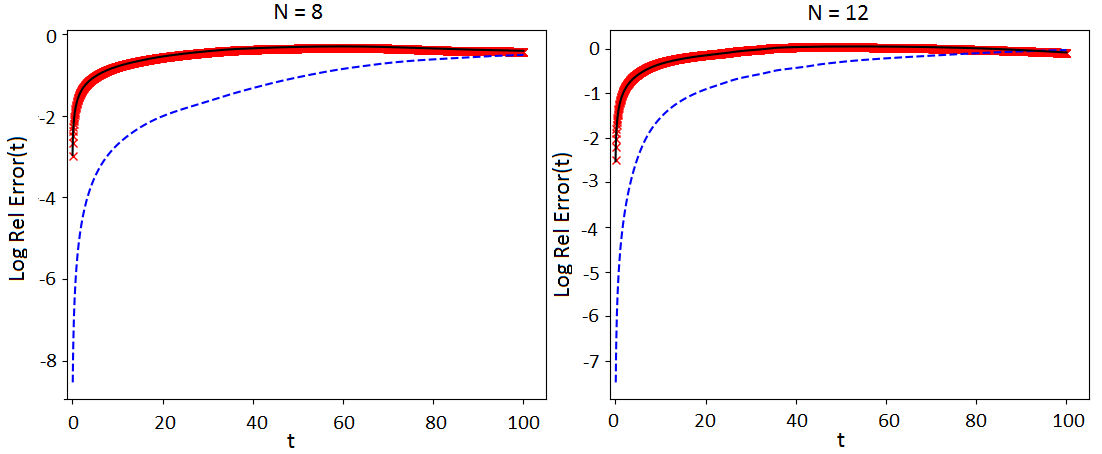}
\end{center}
 \caption{ The logarithmic relative error (R.E.) between the piecewise constant approach and the Magnus series (dashed blue line), the adiabatic approximation (solid black line) and the TDPT approximation (red crosses) against time for the linear case: $\lambda(t) = 10^{-3}t$. These errors have been investigated for different dimensions; $N=2, 4, 8$ and $12$, with a minimum level spacing of 0.01. The Magnus series best approximates $C(t)$, when $t \leq 60$. The accuracy improves with dimension, for $N=12$, the Magnus series best approximates $C(t)$ for $t \leq 100$. This demonstrates that the point of intersection between these R.E.s shift to the right as dimension grows. During the evolution, the R.E.s are bounded by $10^{0}$ for all approximations through time. The R.E. for the Magnus series increases with time as the system approaches a limit such that the convergence criterion in Eq. (\ref{MagConv}) does not hold. The errors for the adiabatic approximation overlaps with the TDPT.}
\label{fig:LinearError}
\end{figure*}

\begin{figure}
\begin{center} 
\includegraphics[width=0.9\linewidth]{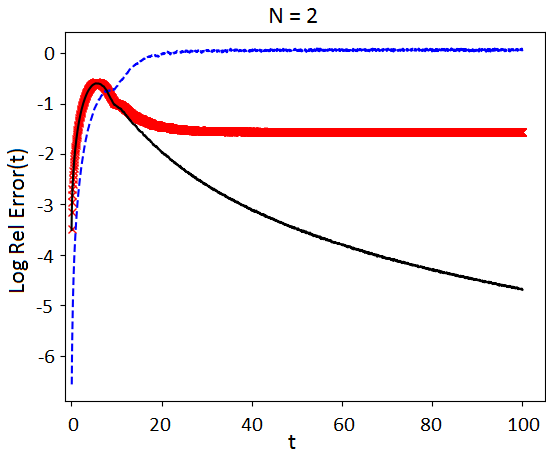}
\includegraphics[width=0.9\linewidth]{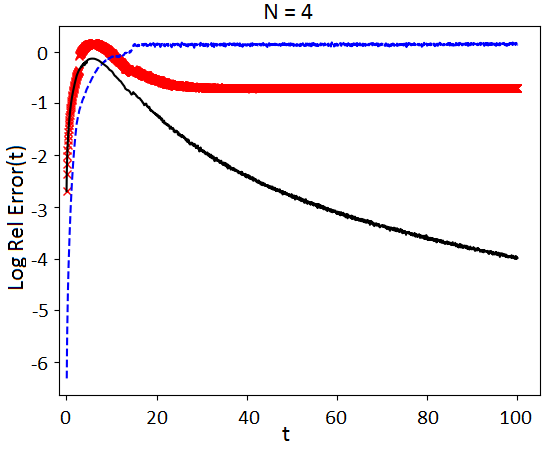}
\includegraphics[width=0.9\linewidth]{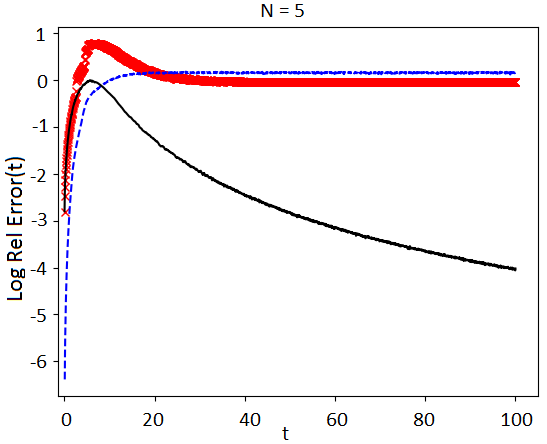}
\end{center}
 \caption{ Same as in Fig. 2, averaged over the same initial conditions, for cubic $ \lambda(t) = 10^{-3}(t^3+t^2+t)$. The errors have been obtained for $N=2, 4$ and $5$, it was not possible to obtain results for larger $N$  as the approximations broke down. The Magnus series best approximates $C(t)$ when $t \leq 10$, and plateaus at $10^{0}$, demonstrating a break down in meeting Eq.(\ref{MagConv}) for the Magnus series. This is expected as the cubic function grows faster than all other classes of $\lambda$ considered in this paper. The relative error for TDPT peaks initially and also plateaus at $10^{0}$, whereas the R.E. for the adiabatic approximation decreases with time. }
\label{fig:LinearError}
\end{figure}

\begin{figure*}
\begin{center} 
\includegraphics[width=0.9\linewidth]{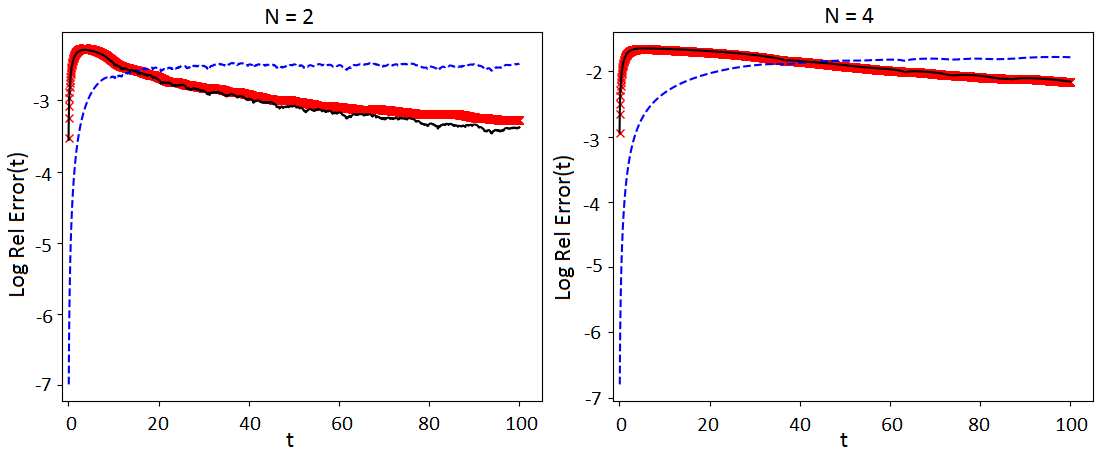}
\includegraphics[width=0.9\linewidth]{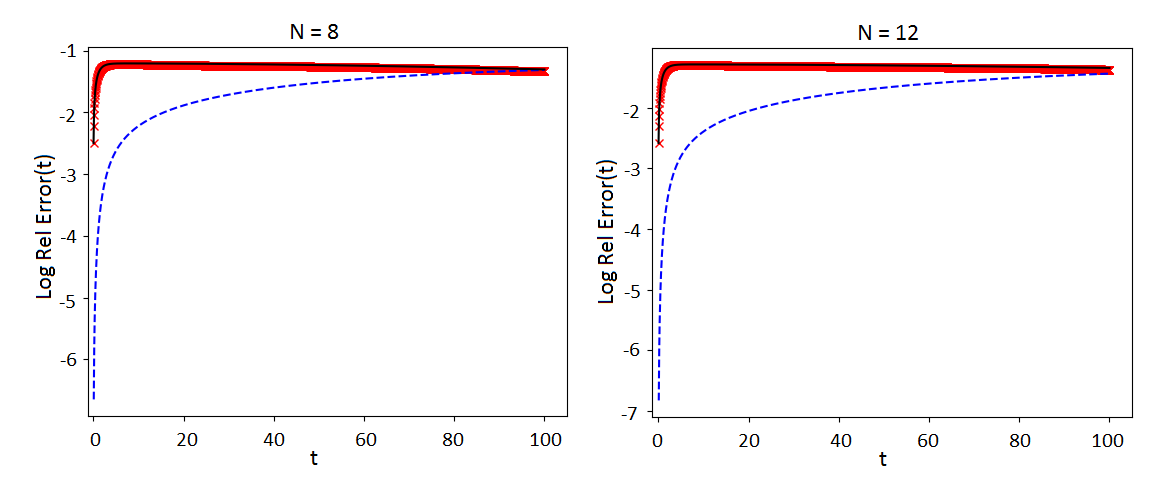}
\end{center}
 \caption{Same as in Fig. 2, again averaged over the same initial conditions, for exponential decay $\lambda(t) = 10^{-3}e^{-t}$. For $t \leq 10$ and $N=2$, the Magnus series best approximates $C(t)$. This period increases with dimension, going beyond $t = 100$ for $N=12$, where the point of intersection between the R.E.s shift to the right as dimension grows. For the exponential decay, the R.E.s for all approximations remain bounded below $10^{-1}$, as time grows large the Magnus series plataues yet provides accurate results throughout the evolution, demonstrating thus far the Magnus series convergence criterion is met. Again, the errors for the adiabatic approximation overlaps with the TDPT, where their errors plateau below $10^{-1}$}
\label{fig:LinearError}
\end{figure*}

\begin{figure}
\begin{center} 
\includegraphics[width=0.9\linewidth]{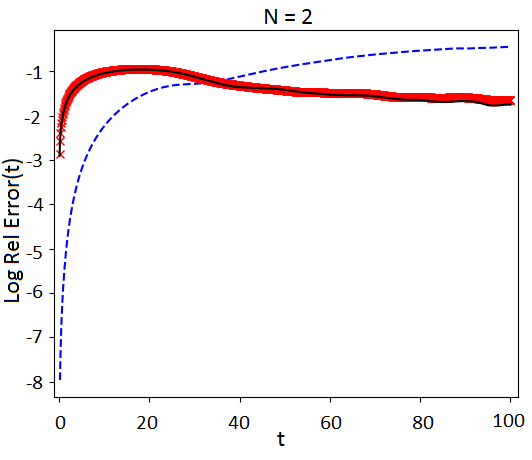}
\includegraphics[width=0.9\linewidth]{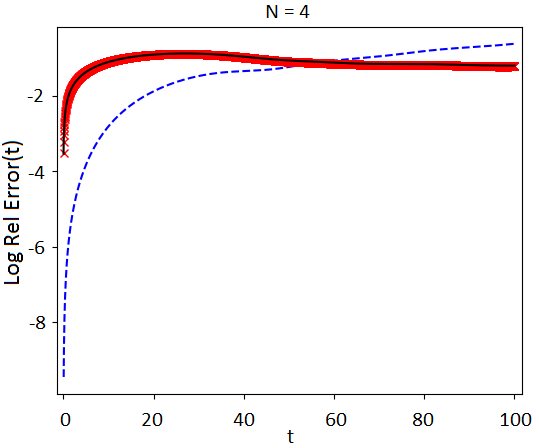}
\includegraphics[width=0.9\linewidth]{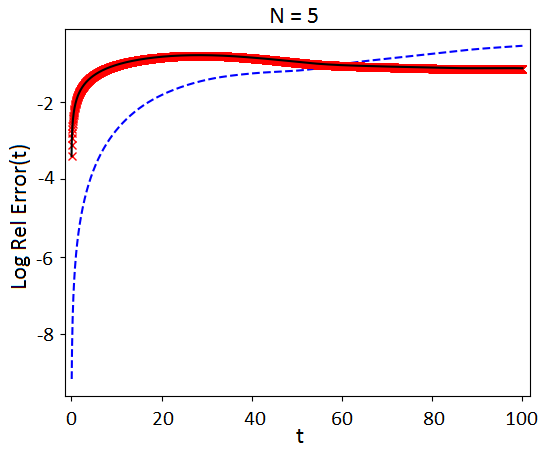}
\end{center}
 \caption{ R.E. between the piecewise constant approach and the Magnus series (dashed blue line), the adiabatic approximation (solid black line) and the TDPT approximation (red crosses) against time for the linear case: $\lambda(t) = 10^{-3}t$. In contrast to Fig. 2, the initial minimum level spacing here is 0.05. These errors have been investigated for dimensions; $N=2, 4$ and $5$. One observes at $N=2$, the Magnus series best approximates $C(t)$ for $t \leq 40$, this period increases with dimension,  at $N=5$, reaching $t \leq 50$. This demonstrates that the point of intersection between these R.E.s shift to the right as dimension grows. During the evolution, the R.E.s are bounded by $10^{0}$ for all approximations. Only for $N=2$ does the Magnus series approach $10^{0}$. There is a growth in R.E with time as the system appraches a limit such that the convergence criterion in Eq. (\ref{MagConv}) does not hold. The errors for the adiabatic approximation overlaps with the TDPT, both appear to decrease as time grows large.}
\label{fig:LinearError}
\end{figure}

\begin{figure}
\begin{center}
\includegraphics[width=0.9\linewidth]{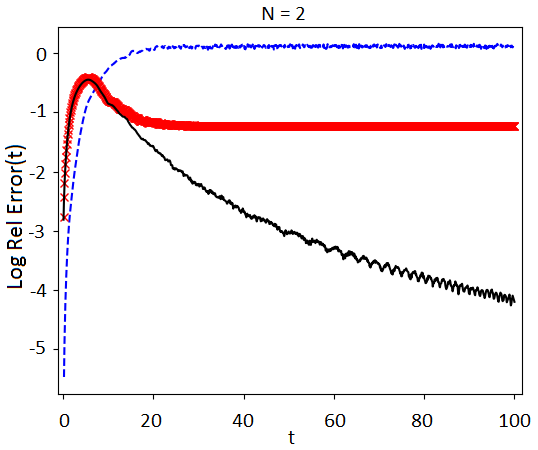}
\includegraphics[width=0.9\linewidth]{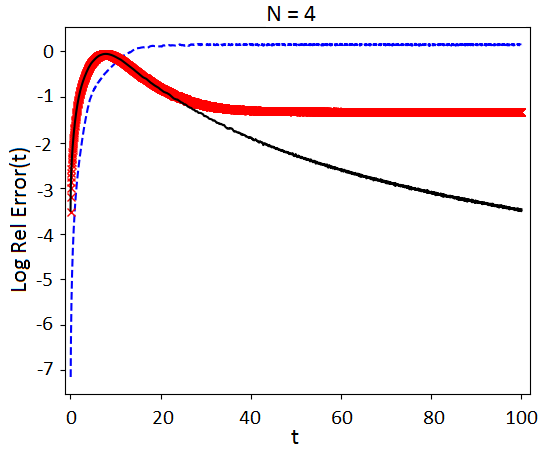}
\includegraphics[width=0.9\linewidth]{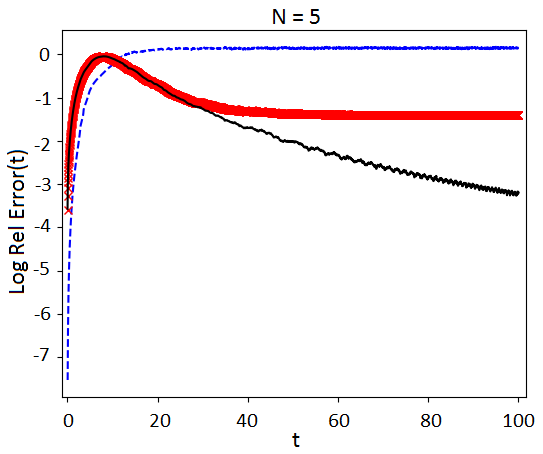}
\end{center}
 \caption{ Same as in Fig. 5, averaged over the same initial conditions, for cubic $ \lambda(t) = 10^{-3}(t^3+t^2+t)$. The Magnus series best approximates $C(t)$ for $t \leq 10$. This interval is shorter than for all other classes of $\lambda$, as the cubic function grows faster than all other classes of $\lambda$ considered in this paper. The R.E. for the Magnus series plateaus at $10^{0}$ for all dimensions, demonstrating a break down in meeting Eq.(\ref{MagConv}). One observes the errors for the adiabatic approximation overlaps with the TDPT. One observes the duration in the overlap increases with dimension however, as time increases the adiabatic approximation is most accurate, decreasing with time, whereas the TDPT plateaus at $10^{-1}$.}
\label{fig:CubicError}
\end{figure}

\begin{figure}
\begin{center}
\includegraphics[width=0.9\linewidth]{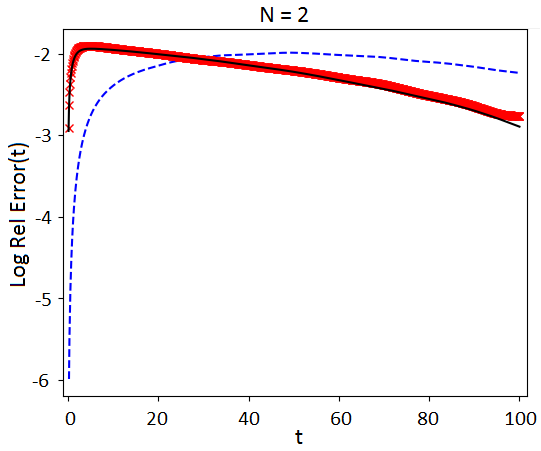}
\includegraphics[width=0.9\linewidth]{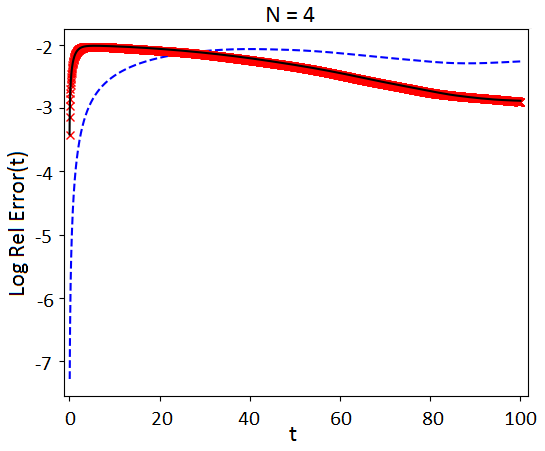}
\includegraphics[width=0.9\linewidth]{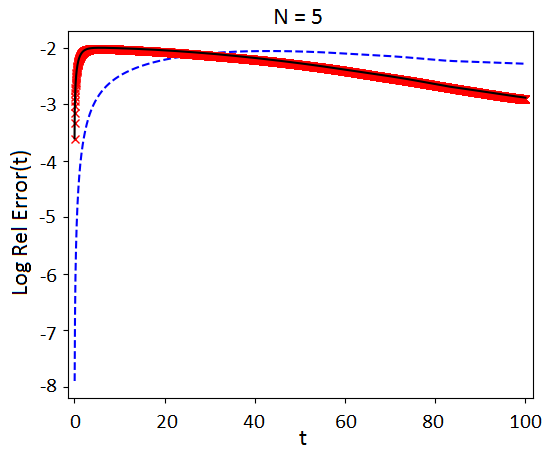}
\end{center}
 \caption{Same as in Fig. 5, again averaged the same initial conditions, for exponential decay $\lambda(t) = 10^{-3}e^{-t}$. For $t \leq 20$, the Magnus series best approximates $C(t)$. This period increases with dimension, reaching $t \leq 30$ at $N=5$, where, again the point of intersection between the R.E.s shift to the right as dimension grows. For the exponential decay, the R.E.s for all approximations remains below $10^{-2}$, as time grows large the Magnus series plataues yet provides accurate results throughout the evolution, demonstrating thus far the Magnus series convergence criterion is met. Again, the errors for the adiabatic approximation overlaps with the TDPT, both seen to decrease as time grows large at the same rate such that beyond $t=30$, these provide better approximations for $C(t)$.}
\label{fig: ExponentialError}
\end{figure}

We observe for short time intervals the best approximation for $C(t)$ is the Magnus series, however as time grows large there is a break down in meeting the convergence critera  Eq.(\ref{MagConv}) for the set initial conditions. Exponential decay is an exception case; the growth of the system is slow enough that the R.E saturates before reaching errors of $10^{-2}$. This provides accurate solutions throughout the evolution. The R.E.s from the adiabatic approximation and the TDPT decrease below the Magnus series R.E. as time grows large, a consequence of the levels becoming further apart resulting in $||P||$ becoming less significant. We note that this weakens the approach. The Magnus series in contrast is well suited to the `spaghetti regime' where levels are close, a result of the Magnus series being less vulnerable to the effects of level crossings (as shown in Appendix B). This is observed in the general trend in Figs. 2, 4, 5 and 7, that for larger N where level interactions are more frequent, the Magnus series relative errors overtake the errors for both the adiabatic and TDPT approximations at later times.  

\section{Exact Cover Algorithm: A Variation on 3-Satisfiability}
To further consider our investigation on the different approximation in Sec. III, we now turn to applying our approach to a concrete example, the exact cover 3-satisfiability problem, comparing the applicabilities of the different approximations to this problem. We determine the eigenstate coefficients from which one obtains the occupation dynamics crucial to the understanding of sources of decoherences in a quantum system. Decoherences arise from a number of various elements intrinsically and from the environment ranging from level (avoided) crossings to random dissipative influences from the environment, however the investigation of these various sources are beyond the scope of this paper.

The exact cover algorithm, belongs to the class of NP-complete problems\cite{Knuth, Wang}, first proposed by Knuth\cite{Knuth}. It has since been extended to the AQC setting\cite{Fahri1}, cast as a variation on 3-satisfiability\cite{Fahri1, Fahri2, Wang}. The problem is described by a Boolean expression, the intersection of all clauses for a string of $N$ binary variables in a set $S$, constrained by $M$ clauses, each acting on three variables; $y_{\alpha}, y_{\beta}$ and $y_{\gamma}$ with $\alpha, \beta, \gamma \in \mathbb{N}$. The clause is satisfied if and only if one of the three variables takes the value 1 whilst the other two take 0; $y_{\alpha}+y_{\beta}+y_{\gamma}=1$, described by the clause function such that each violated clause is associated with a fixed energy penalty\cite{Wang}: $\sum_{Clauses}(y_{\alpha}+y_{\beta}+y_{\gamma}-1)^2$ used to obtain a solution to the problem. The Hamiltonian describing this problem can be translated to an $M$-qubit problem, given by the following: 


\begin{equation}
\begin{gathered}
\label{ExactCover}
H=\lambda\sum^M_{i=1}\frac{1-\sigma^x_i}{2}+(1-\lambda)\sum^M_{i<j}C_{ij}(1-\sigma^z_i)(1-\sigma^z_j), 
\end{gathered}
\end{equation}

\noindent where $C_{ij} \in \mathbb{N}$ counts the pairwise occurrence of any two distinct variables in the clauses and $\sigma^x$ and $\sigma^z$ are given by the Pauli spin matrices, translating the description through qubits. 

We consider three distinct clauses with $C_{12}=C_{23}=2$ and $C_{13}=1$ with an exponential decay function for $\lambda = 10^{-3}e^{-t}$. The energy spectrum is determined by diagonalising Eq.(\ref{ExactCover}), giving the eigenvalues. Combined with Eq.(\ref{Pechukas}) we determine the evolution of the level dynamics. We note here that the initial conditions do not meet Eq.(\ref{MagConv}). However, as one observes in Figs. 2, 4, 5 and 7, the Magnus series is robust in that despite the initial conditions having satisfied the criterion outlined in Eq. (\ref{MagConv}) for in the interval [0, 1], the approximation had accurately provided solutions far beyond this duration. Using the flexibility observed in Figs. 2, 4, we compare the different approximations explored in Sec. III to obtain the evolution of the eigenstate coefficients, up to $t=100$ in steps of 0.01 with Gaussian distributed initial conditions, normalised for $C(0)$. We determine the logarithm of the relative errors compared against the piecewise constant approach for each approximation through time in Fig. 5.  

\begin{figure*}
\begin{center}
\includegraphics[width=0.9\linewidth]{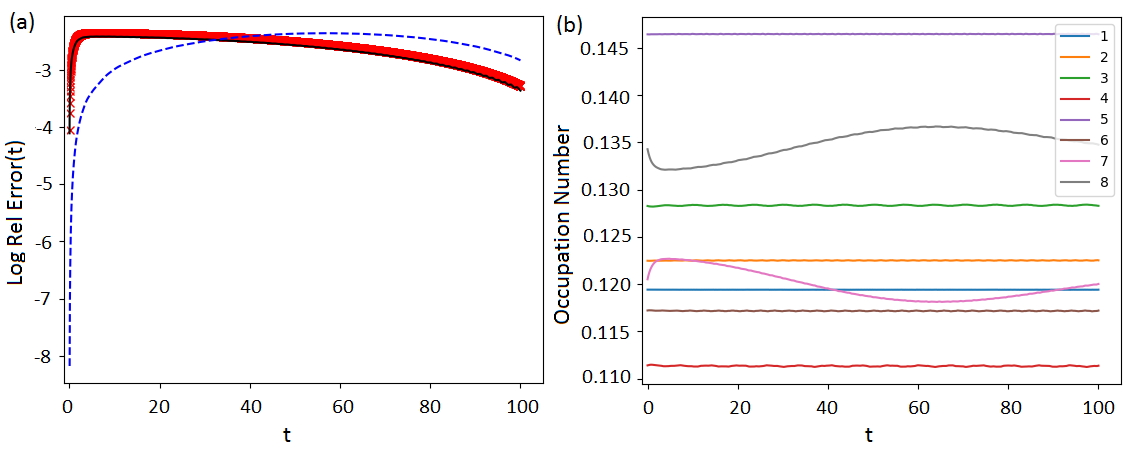}
\end{center}
 \caption{ (a) The logarithm of the relative error against time for the adiabatic approximation (thick crosses), the TDPT (solid line) and the Magnus series approximation (dashed line). One observes the errors throughout the evolution, in all cases are bounded by $10^{-2}$. The adiabatic approximation overlaps with the TDPT, however up to $t \leq 35$, the Magnus series best approximates $C(t)$ providing accurately the dynamics of the eigenstate coefficients. (b)  The evolution of the occupation numbers of the 3-Satisfiablity qubit system to study the exact cover 3 problem of 8 bits. We observe the presence of an avoided crossing between states 7 and 8, resulting in a reflection in their occupation dynamics, suggesting a transfer in the population of states. In contrast, all other states have remained essentially constant despite a level crossing between states 7, 1 and 2. It would be of interest to determine the dynamics under the influence of noise modelling interactions with the environment.}
\label{fig:ExactCover}
\end{figure*}

We observe the Magnus series best approximates the evolution of the eigenstate coefficients throughout the duration, with the error bounded below $10^{-2}$ up to $t \leq 35$. Using the relation $\rho=C(t)\otimes C^{*T}(t)$, where $C^{*T}(t)$ denotes the complex conjugate transpose of $C(t)$, we determine the evolution of the density matrix for this system hence we obtain the dynamics of the occupation numbers, given in Fig. 6.

One obtains the evolution of the occupation numbers from diagonalising the density matrices, these describe the probability of remaining in the initial states where the off-diagonal terms describe the dynamics of the coherences, giving the probability of state transitions. Using this description, one can explore various sources of decoherence from stochastic processes as well as Landau-Zener transitions from interactions between the levels and their impact on the population of states.  

These investigations on the evolution of eigenstate coefficients could be realised experimentally. For example, consider the experiments by D-Wave One concerning 108 qubits\cite{Boixo, War}. One could translate their quantum annealed Hamiltonian based on the Ising model to the Pechukas-Yukawa setting using Eq. (\ref{Pechukas}); choosing some function of $\lambda(t)$ to satisfy the start and end points of an interval such that it simulates time over $t \in [0, t_f]$\cite{Boixo, War}. Under this description, one could then approximate the eigenstate coefficients which can then be used to determine the occupation numbers and coherences as the system evolves in time. Similarly, one could apply this analysis to the D-WaveTwo experiments, regarding 512 qubits which held experimentally for  $t \in [0, t_f]$ where $t_f=20\mu s$. 

\section{Discussion and Conclusions.}
\noindent In this paper we have investigated the relation between the level dynamics and the evolution of the quantum states under a Pechukas formalism. Three different approaches were taken to approximate the eigenstate coefficients: a Magnus series expansion, an adiabatic approximation and time dependent perturbation theory.  Numerically, it was found that for short time intervals, where the convergence criterion is satisfied, the Magnus series was most accurate. In these intervals, the R.Es for the Magnus series was lower by multiple orders than both the adiabatic and TDPT approximations. We further investigated the limits of the Magnus series, reducing the convergence criterion such that the entire evolution of the system is governed by the initial conditions and choice in $\lambda$. We found that the Magnus series is robust, in that as time evolves the error increases yet remains the better approximation much beyond the interval where the convergence criterion is satisfied. 

It was observed that the R.E. for the Magnus series had overtaken the R.E. for both the Adiabatic approximation and the TDPT at later times as the number of excited states increased. This demonstrates that the Magnus series is better suited to the "spaghetti" regime, less prone to divergences in the error for level (avoided) crossings which becomes more prevalent for larger $N$. In contrast, particularly the TDPT is sensitive to these level interactions resulting in the perturbations becoming large. Both the adiabatic and the TDPT approximations errors overlaped for the linear and exponential $\lambda$ evolutions. Only during cubic evolutions for $\lambda$ did the adiabatic approximation hold better than both Magnus and TDPT approximations. Our work on the Magnus series against the adiabatic approximation, under theoretically similar parameters compares similarly with that by Pan et al in [\onlinecite{Pan}]. Under different settings, our work comparing the Magnus series against the TDPT also compares well with Blanes et al in [\onlinecite{Blanes}].

Under this description, one gains the potential to explore the different classes of Hamiltonians in the Pechukas-Yukawa formalism in order to determine how they may be solved based on the initial conditions and the complexity classes they fall under. Under the Magnus series description of $C(t)$, one can derive from the bounds given in Eq.(\ref{ConvPechX}) and Eq.(\ref{ConvPechP}), the set of initial conditions to satisfy the convergence criterion outlined in Eq. (\ref{ConvPech}) over a desired duration. This could yield relevence in specifying Hamiltonians of different complexity classes.


The Magnus series provides an infinite hierarchy in powers of $\dot\lambda$ parameters. The structure is that of a cumulant expansion and it would be of interest to consider asymptotic convergences in the Magnus series, improving the efficiency of the result. Using such an expansion it will be interesting to consider $C(t)$ in the adiabatic limit, as $\dot\lambda$ goes to $0$. One may explore the significance of these terms with respect to the developments of adiabatic invariants. This has the potential to significantly impact features of the adiabatic algorithm design. Furthermore, this description enables the exploration of the relationship between the level dynamics and that of the dynamics of the quantum states described by the evolution of the density matrix, developing on the model established in [\onlinecite{Ours}]. This would provide analytical insight into the sources of decoherence on the evolution of a quantum system. These range from the effects of noise due to interactions with the environment; leading to dissipative influences on state populations, to intrinsic sources; a consequence of level (avoided) crossings, currently modelled by Landau-Zener transitions. Under the description of the density matrix, one obtains a more detailed picture of the dynamics of the populaion of states independent of the simplifications imposed by the Landau-Zener model; these shall be explored further in future investigations.

\section*{Acknowledgments} We are grateful to Sergey Savel'ev, Alexander Veselov, Anatoly Nieshtadt, Alec Maassen van den Brink and Patrick Navez for the valuable discussions that greatly improved the manuscript. This work has been supported by EPSRC through the grant No. EP/M006581/1.

\section*{Appendix A: Magnus Criterion-First Integral}

Rewriting the $||X||$ integral in Eq.(\ref{ConvPech}) in terms of initial conditions $x_n(0), v_n(0),  l_{mn}(0)$, we consider the Lax formalism in order to express the Pechukas equations  Eq.(\ref{Pechukas}) by \cite{Light, YukawaLax, Chowski}:

\begin{equation}
\begin{gathered}
\label{PechukasLax}
\dot{X}=W+[P,X] \\
\dot{W}=[P,W]\\
\dot{L}=[P,L],
\end{gathered}
\end{equation}

\noindent where $P$ is as expressed in Eq.(\ref{EvolCMatrix}),  matrices $W$ and $L$ are skew-Hermitian, given by:

\[W=w_{mn}$ where $w_{mn}=\frac{l_{mn}}{\left(x_m-x_n\right)}$ and $w_{mm}=0\]
\[L=l_{mn}$ and $l_{mm}=0.\]

\noindent As before, $X=\mathrm{diag}\left(x_1\dots x_n\right)$ denotes the diagonal matrix of the eigenvalues of the system. $X$ can be transformed,  through a unitary transformation to a nondiagonal matrix $Y$, $X=UYU^{-1}$, 
where $U$ is a matrix of eigenvectors. The matrix $Q$ is defined by:
\[Q=W+\mathrm{diag}\left(v_1\dots v_n\right).\]

%
%
In Lax formalism, $Y$ is then expressed in terms of the initial conditions\cite{Chowski}:

\begin{equation}
\begin{gathered}
\label{YInitialCond}
Y(t)=\lambda(t)Q(0)+X(0).
\end{gathered}
\end{equation}

\noindent Time dependence exists solely through the evolution of $\lambda$. Using the unitary transformation of $X$ 
and Eq.(\ref{YInitialCond}), then  $||X(t)||=||Y(t)||=\sqrt{Tr(Y^*(t)Y(t))} =\sqrt{||X(0)||^2+\lambda(t)Tr(X(0)Q(0))+\lambda^2(t)||Q(0)||^2}$. 
Substituting this for the $||X||$ integral in Eq.(\ref{ConvPech}), we obtain:

\begin{widetext}
\begin{equation}
\begin{gathered}
\label{ConvPechXApp}
\int^t_{0}{\sqrt{||X(0)||^2+\lambda(s)Tr(X(0)Q(0))+\lambda^2(s)||Q(0)||^2}ds} \\
\leq t||X(0)||+\sqrt{Tr(X(0)Q(0))}\int_0^t{\sqrt{\lambda(s)}ds}+||Q(0)||\int_0^t{|\lambda(s)|ds}.
\end{gathered}
\end{equation}
\end{widetext}

\noindent We reduced the convergence of the $X$ integral solely to the dependence of initial conditions and the time evolution of $\lambda$, taking advantage of the Pechukas dynamics being encoded by the initial conditions and that they are expressible in Lax formalism. 


\section*{Appendix B: Magnus Criterion-Second Integral}

Level crossings may result in Landau-Zener transitions of the population of states. These occur at a $\lambda^*$, potentially involving multiple levels which is considered separately. Note that in the $N=2$ case described by the Landau-Zener model, the system collapses to the Calegro-Sutherland model with constant $l_{mn}$ terms. We show in this section that level crossings due to the symmetries of the Hamiltonian, have zero measure. 

For a level crossing $x_m=x_n$ at $\lambda^*$ , then Eq.(\ref{Pechukas}) implies $l_{mn}=0$ and $\dot l_{mn}=0$. The converse is not necessarily true, that is if $l_{mn}=0$ does not imply $x_m=x_n$. Expanding about this point with $\delta\lambda^*=(\lambda-\lambda^*)$, we obtain the following expression for the upper bound on Eq. (\ref{ConvPechP}): 

\begin{widetext}
\begin{equation}
\begin{gathered}
\label{ExplicitExpandP}
\sum^N_{m\neq n}\frac{|l_{mn}(\lambda^*)+\delta\lambda\dot l_{mn}(\lambda^*)+\frac{1}{2}\delta\lambda^2\ddot l_{mn}|}{(x_m(\lambda^*)-x_n(\lambda^*))^2+2\delta\lambda(x_m(\lambda^*)-x_n(\lambda^*))(v_m(\lambda^*)-v_n(\lambda^*))+\delta\lambda^2(v_m(\lambda^*)-v_n(\lambda^*))^2}+\mathcal{O}(\lambda^3),
\end{gathered}
\end{equation}
\end{widetext}

\noindent where $\ddot l_{mn}$ is given by $\sum^N_{k\neq m,n}\frac{-2l_{mk}l_{kn}(v_m-v_n)}{(x_m-x_k)^3}$. Cancelling zero valued terms and substituting $\ddot l_{mn}$ into Eq. (\ref{ExplicitExpandP}),





\begin{equation}
\begin{gathered}
\label{ExpandP}
\frac{|l_{mn}|}{(x_m-x_n)^2}=\left. \sum^N_{k \neq m,n}\frac{|l_{mk}l_{kn}+\mathcal{O}(\lambda^3)|}{(x_m-x_k)^3(v_m-v_n)+\mathcal{O}(\lambda^3)}\right\vert_{\lambda^*}.
\end{gathered}
\end{equation}

This series diverges in two scenarios, case 1: degenerate level crossings: $x_k = x_m = x_n$ for some $k$, which again by  Eq.(\ref{Pechukas})  gives $l_{mn}, l_{mk}, l_{nk}$ vanishes and case 2: that $v_m=v_n$ describing a system where levels coalesce.
For case 1, as both numerator and denominator are zero, warrants the application of l'Hopital's rule on Eq.(\ref{ExpandP}). At it's third iteration, we obtain:  

\begin{equation}
\begin{gathered}
\label{L'hopital}
\frac{|l_{mn}|}{(x_m-x_n)^2}= \sum_{k \neq m,n}\frac{0+\mathcal{O}(\lambda^3)}{6(v_m-v_n)(v_m-v_k)^3+\mathcal{O}(\lambda^3)}.
\end{gathered}
\end{equation}

\noindent The expression converges to zero at the critical point $\lambda^*$, implying that degenerate level crossing do not cause Eq.(\ref{ExpandP}) to diverge.






Exploring case 2, we use the interpretation of the Pechukas equations as describing a 1D gas. As $\lambda$ approaches $\lambda^*$; $\lambda^{-}=  \lambda^*-\epsilon$ and without loss of generality $x^{-}_m>x^{-}_n$, it is clear that $v_m = \lim_{\epsilon \rightarrow 0} \frac{x_m-x^{-}_m}{\epsilon}$ hence $(v_m-v_n) \approx \frac{(x^{-}_m-x^{-}_n)}{\epsilon}$ greater than 0 by assumption. By symmetry, this argument holds for $x^{-}_n > x^{-}_m$.
In the case $v_m=v_n$, at $\lambda^*$ we consider the difference between acceleration terms given by the following:

\begin{equation}
\begin{gathered}
\label{Acceleration}
\frac{(\dot v_m-\dot v_n)}{2} =
 \sum^N_{k \neq m,n}\left(\frac{|l_{mk}|^2}{(x_m-x_k)^3})-\frac{|l_{nk}|^2}{(x_n-x_k)^3}\right) \\+\frac{|l_{mn}|^2+|l_{nm}|^2}{(x_m-x_n)^3}.
\end{gathered}
\end{equation}

\noindent The latter term corresponding to the level crossing, tends to $0$ as $\lambda \rightarrow \lambda^*$ as determined by the application of l'Hopitals rule three times, however the terms in the sum are non-zero, describing acceleration between the levels at $\lambda^*$, modelling repulsion such that levels do not coalesce. This shows that level crossings occur only for an instant $\lambda^*$ rather than intervals, as such they do not contribute to Eq.(\ref{ConvPechP}) as they have zero measure. 


\end{document}